\documentclass[twocolumn,secnumarabic,amsmath,amssymb, nobibnotes, aps, prb,groupedaddress,superscriptaddress]{revtex4-2}
\usepackage{amsmath,graphicx,latexsym,times,color}
\usepackage{setspace}
\usepackage{hyperref}
\usepackage{array}
\usepackage{textcomp}
\usepackage{titlesec}
\usepackage{physics}
\usepackage{gensymb}
\usepackage{nccmath}
\usepackage{empheq} 
\usepackage{fontawesome5}
\usepackage{enumerate}
\usepackage{textcomp}
\usepackage{gensymb}
\usepackage[cmyk,dvipsnames]{xcolor}
\usepackage[utf8x]{inputenc} 
\usepackage{comment}
\usepackage[normalem]{ulem}    
\usepackage[dvipsnames]{xcolor} 

\definecolor{alizarin}{rgb}{0.82, 0.1, 0.26}

\usepackage{lettrine}

\let\oldtimes\times  
\renewcommand\times{{\oldtimes}}

\newcommand\redsout{\bgroup\markoverwith{\textcolor{red}{\rule[0.5ex]{2pt}{0.4pt}}}\ULon}

\renewcommand{\vec}[1]{\mathbf{#1}}

\usepackage[normalem]{ulem}

\definecolor{darkorchid}{HTML}{bf3eff}

\definecolor{dsgrey}{rgb}{0,1,1}

\newcommand{\remove}[1]{\bgroup\markoverwith{\textcolor{red}{\rule[0.5ex]{2pt}{0.4pt}}}\ULon{#1}\egroup}

\begin{document}

\title{Sub-20 nm spin textures with arbitrary topological charge stabilized by higher-order interactions}

\author{Zhen Sun}
\affiliation{Zhejiang Key Laboratory of Quantum State Control and Optical Field Manipulation and Department of Physics,
\href{https://ror.org/03893we55}{Zhejiang Sci-Tech University}, Hangzhou 310018, China}
\affiliation{\href{https://ror.org/01ahyrz84}{Universit\'e de Toulouse}, \href{https://ror.org/02feahw73}{CNRS}, \href{https://ror.org/03kwnqq69}{CEMES}, Toulouse, France}

\author{Shiwei Zhu}
\affiliation{\href{https://ror.org/01ahyrz84}{Universit\'e de Toulouse}, \href{https://ror.org/02feahw73}{CNRS}, \href{https://ror.org/03kwnqq69}{CEMES}, Toulouse, France}

\author{Moritz A. Goerzen}
\affiliation{\href{https://ror.org/01ahyrz84}{Universit\'e de Toulouse}, \href{https://ror.org/02feahw73}{CNRS}, \href{https://ror.org/03kwnqq69}{CEMES}, Toulouse, France}

\author{Megha Arya}
\affiliation{\href{https://ror.org/01ahyrz84}{Universit\'e de Toulouse}, \href{https://ror.org/02feahw73}{CNRS}, \href{https://ror.org/03kwnqq69}{CEMES}, Toulouse, France}

\author{Changsheng Song}
\email[\faEnvelope\ e-mail: ]{cssong@zstu.edu.cn}
\affiliation{Zhejiang Key Laboratory of Quantum State Control and Optical Field Manipulation and Department of Physics,
\href{https://ror.org/03893we55}{Zhejiang Sci-Tech University}, Hangzhou 310018, China}

\author{Dongzhe Li}
\email[\faEnvelope\ e-mail: ]{dongzhe.li@cemes.fr}
\affiliation{\href{https://ror.org/01ahyrz84}{Universit\'e de Toulouse}, \href{https://ror.org/02feahw73}{CNRS}, \href{https://ror.org/03kwnqq69}{CEMES}, Toulouse, France} 
	\date{\today}
	
\begin{abstract}

\noindent 

Topological magnetization textures play a central role in modern magnetism. For many applications, spin textures with different topological charges $Q$ in the same system are particularly attractive.
Although the coexistence of skyrmions and antiskyrmions ($|Q| = 1$) has been reported in inversion-symmetric magnets, extending this to high-$Q$ ($|Q| > 1$) textures with arbitrary charge remains elusive. Here, using an atomistic spin model parameterized from first-principles calculations, we predict the emergence of sub-20 nm high-$Q$ textures in Janus monolayers, van der Waals magnets of growing interest. We explore skyrmion and antiskyrmion rings as well as skyrmion bags with $|Q|$ up to 5, and characterize their nucleation mechanisms, thermal stability, and collapse pathways. We find that higher-order spin interactions (HOI), which extend the conventional bilinear-exchange Hamiltonian, are essential for stabilizing these high-$Q$ spin textures. The rings remain thermally stable at zero magnetic field. HOI substantially enhance their energy barriers while leaving their size nearly unchanged and stabilize them even in the absence of Dzyaloshinskii-Moriya interaction. Skyrmion bags, in contrast, nucleate only in the presence of HOI. In particular, the four-spin three-site interaction is the key ingredient preventing high-$Q$ textures from collapsing into the ferromagnetic state. Finally, we identify previously unreported parity-dependent collapse mechanisms for high-$Q$ textures. Our results establish HOI as an overlooked mechanism for high-$Q$ nucleation.

\emph{}\\

\noindent DOI: $\times \times \times$ \hfill Subject Areas: Condensed Matter Physics, Magnetism, Spintronics
\end{abstract}
	
\maketitle

\section{INTRODUCTION}

Magnetic topological solitons are localized, vortex-like spin textures characterized by a nontrivial integer topological charge $Q$ \cite{muhlbauer2009skyrmion,fert2017magnetic,gobel2021beyond}, defined as
\begin{equation}\label{topo_charge}
Q=\frac{1}{4\pi}\int_{\mathbb{R}^2}\mathbf{m}\cdot
\left(
\frac{\partial\mathbf{m}}{\partial x_1}\times
\frac{\partial\mathbf{m}}{\partial x_2}
\right)
\,\mathrm{d}^2\mathbf{r},
\end{equation}
where $\mathbf{m}(\mathbf{r})$ denotes the normalized local magnetization. Solitons with different topological charges are usually separated by finite energy barriers because changing $Q$ requires crossing a singular spin configuration, giving rise to topological protection \cite{polyakov22metastable}.

Magnetic skyrmions were originally discovered in a series of non-centrosymmetric compounds, where Dzyaloshinskii-Moriya interaction (DMI) is the key ingredient for their stabilization. Most such systems host only skyrmions ($Q=-1$). High-$Q$ ($|Q|>1$) spin textures, such as skyrmion bundles and skyrmion bags, have recently been reported theoretically \cite{PRL2017,Rybakov2019,Kuchkin2020,Zhouyan2016} and experimentally \cite{Foster2019,Tang2021,hassan2024dipolar,Niu2025,wu2026current}. However, they typically have lateral dimensions ranging from hundreds of nanometers to micrometers. For practical spintronic applications, high-$Q$ states must combine thermal stability close to room temperature, diameters below 20 nm, and efficient control by external stimuli (e.g., strain, electric fields, or light). Existing materials meet some of these requirements, but none meet all three.

\begin{figure*}[!tbp]
	\centering
	\includegraphics[width=1\linewidth]{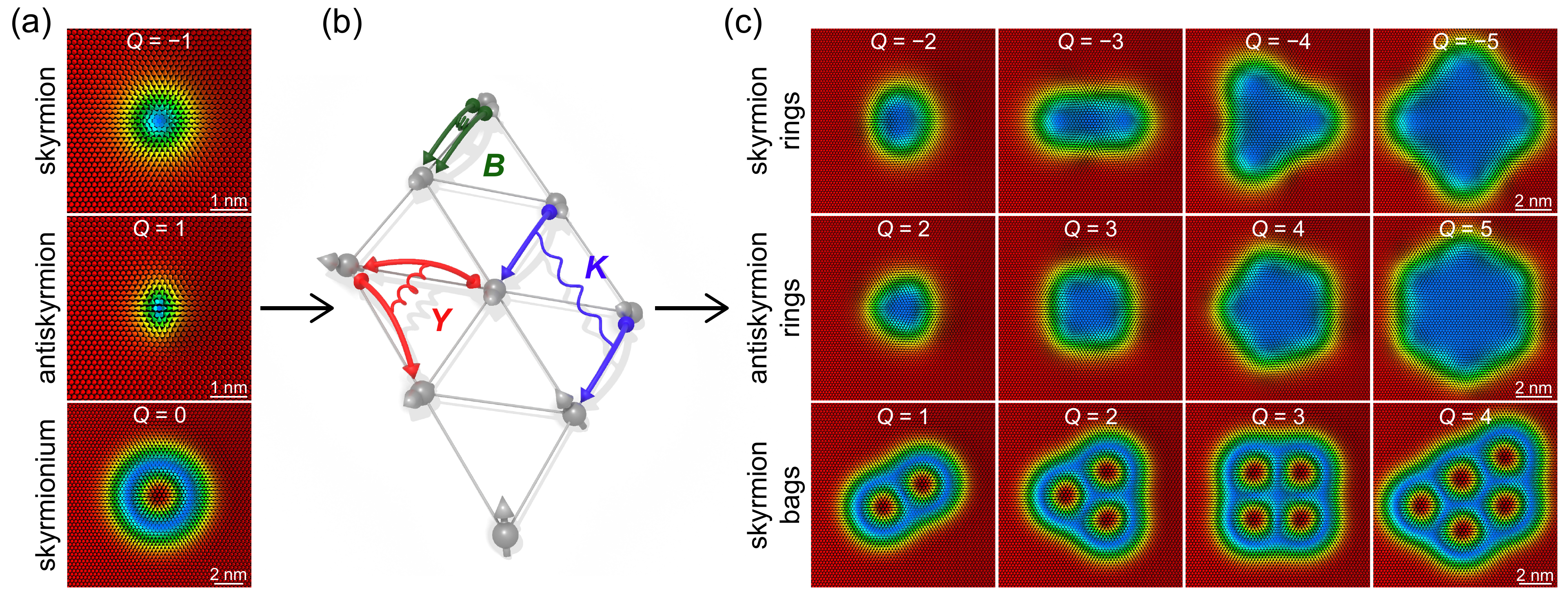}
	\caption{
    Emergence of high-$Q$ topological spin textures driven by higher-order interactions (HOI).
    (a)~Basic spin texture units consisting of a skyrmion ($Q=-1$), an antiskyrmion ($Q=1$), and a skyrmionium ($Q=0$).
    (b)~Schematic representation of HOI on a triangular lattice. Here, HOI couple two of the exchange interactions to form four-spin interactions involving two ($B$, green), three ($Y$, red), or four ($K$, blue) sites as indicated by the springs.
    (c)~High-$Q$ topological spin textures obtained with HOI, comprising skyrmion rings ($Q=-2$ to $-5$), antiskyrmion rings ($Q=2$ to $5$), and skyrmion bags ($Q=1$ to $4$). All spin textures are on a sub-20 nm length scale. Every texture shown is fully relaxed with the fourth-order interactions included.
    }
    \label{overview}
\end{figure*}

The search has therefore shifted to realistic spin models for beyond-Heisenberg solids. Higher-order exchange interactions (HOI) beyond the pairwise Heisenberg exchange naturally arise from the Hubbard model, including the four-spin two-site  (biquadratic) and four-spin four-site interactions. The biquadratic interaction has been shown to renormalize magnetic properties \cite{kartsev2020biquadratic} and be essential for describing magnetic ground states in 2D van der Waals (vdW) magnets \cite{Xiang2021}. The interplay between Heisenberg exchange and the four-spin four-site interaction was shown to stabilize atomic-scale skyrmion lattices, in excellent agreement with atomic resolution spin-polarized scanning tunneling microscopy (SP-STM) measurements \cite{heinze2011spontaneous}. More recently, Hoffmann \textit{et al.} \cite{Hoffmann2020} derived the four-spin three-site interaction based on a multi-band Hubbard model for spin systems with $S\ge1$. Since Mn- and Cr-based magnets typically host local spin moments on the order of $2$ or $3\,\mu_{\mathrm{B}}$, they provide a natural materials platform for realizing large four-spin three-site interactions. Although HOI are now recognized as an important ingredient for stabilizing low-$Q$ topological spin textures \cite{paul2020role,gutzeit2022nano,Weiyi2024,megha2025}, their role in stabilizing high-$Q$ textures remains largely unexplored. In particular, it is unknown whether HOI can stabilize states with arbitrary $Q$, how such states collapse, and how HOI affect the corresponding energy barriers.

In this work, we propose a novel stabilization mechanism driven by HOI for high-$Q$ spin textures, including skyrmion/antiskyrmion rings and skyrmion bags, in vdW magnets.
Our predictions are based on an atomistic spin model parameterized by density functional theory (DFT) calculations. Our chosen system is the Janus monolayer MnPCl, a representative member of the rapidly emerging class of 2D vdW magnets. We show that the four-spin three-site interaction stabilizes the coexistence of high-$Q$ skyrmion and antiskyrmion rings, even in the absence of DMI, as well as skyrmion bags with arbitrary $Q$. These high-$Q$ states combine sub-20 nm diameters with high stability, thereby overcoming the conventional size--stability limitation. We further show that the four-spin three-site interaction strongly modifies the saddle point (SP) configuration and is mainly responsible for the large increase in the collapse barriers. Finally, we discover previously unknown collapse mechanisms unique to high-$Q$ topological spin textures. Our work also opens new perspectives for enhancing the stability of topological spin structures -- even in systems with inversion symmetry, \textit{i.e.}, in the absence of the DMI. 

\begin{table*}[!t]
    \centering
    \scalebox{1.0}{
    \begin{tabular}{ccccccccccccc}
    \hline\hline
    & ~~~~lattice~~~~ & ~~~~$J_1$ ~~~~& ~~~~$J_2$~~~~ & ~~~~$J_3$~~~~ & ~~~~$D_1$~~~~ & ~~~~$D_2$~~~~ & ~~~~$D_3$~~~~ & ~~~~$K_u$~~~~ & ~~~~$B_1$~~~~ & ~~~~$Y_1$~~~~ & ~~~~$K_1$~~~~ &~~~~$\mu$~~~~  \\ 
    \hline
    MnPCl & ~~~~triangle ~~~~ & 32.51 & 4.37 & $-8.66$ & $-0.88$ & 0.04 & $-0.01$ & 0.05 & 1.70 & $-3.13$ & 0.04 & 3.344\\
	\hline
    \end{tabular}}
        \caption{Magnetic interactions including HOI for MnPCl monolayer.
    Shell-resolved exchange constants ($J_i$), DMI constants ($D_i$), MAE ($K_u$), biquadratic ($B_1$), four-spin three-site ($Y_1$), and four-spin four-site ($K_1$) constants for monolayer MnPCl. All values are given in meV/atom, except for the spin moment $\mu$, which is expressed in $\mu_{\text{B}}$/atom. All interaction parameters are taken from the DFT calculations of Ref.~\cite{Hongxin2023} and converted to the convention of Eq.~(\ref{model}), as detailed in Supplementary Note~1 of SM \cite{supplmat}.
    }\label{magnetic_parameters}
\end{table*}

\section{THEORETICAL METHODS}
\label{theory}

\subsection*{A. First-principles atomistic spin model}

To describe the magnetic state of 2D vdW magnets, we employ an atomistic spin model with classical vector spins $\vec{m}_i$ of unit length placed on atomic sites $i, j, k, l$ consisting of the following Hamiltonian of the beyond-extended Heisenberg model:

\begin{equation}\label{model}
\begin{split}
H = &-\sum_{ij}J_{ij}(\vec{m}_i\cdot\vec{m}_j)
-\sum_{ij}\vec{D}_{ij}\cdot(\vec{m}_i\times\vec{m}_j) \\
&-K_u\sum_i (m_i^z)^2
-\mu B\sum_i m_i^z \\
&-B_1\sum_{\langle ij\rangle}
(\vec{m}_i\cdot\vec{m}_j)^2 \\
&-2Y_1\sum_{\langle ijk\rangle}
(\vec{m}_i\cdot\vec{m}_j)
(\vec{m}_j\cdot\vec{m}_k) \\
&-K_1\sum_{\langle ijkl\rangle}
\Big[
(\vec{m}_i\cdot\vec{m}_j)
(\vec{m}_k\cdot\vec{m}_l)
\\
&\qquad +
(\vec{m}_i\cdot\vec{m}_l)
(\vec{m}_j\cdot\vec{m}_k)
-
(\vec{m}_i\cdot\vec{m}_k)
(\vec{m}_j\cdot\vec{m}_l)
\Big].
\end{split}
\end{equation}
The first four magnetic interaction terms in Eq. (\ref{model}) correspond to Heisenberg exchange ($J_{ij}$), the DMI vector ($\vec{D}_{ij}$), the magnetocrystalline anisotropy energy (MAE, $K_u$), and the Zeeman coupling to an external field ($B$), where $\mu$ is the magnetic moment. The last three are the fourth-order HOI, namely the biquadratic ($B_1$), four-spin three-site ($Y_1$), and four-spin four-site ($K_1$) terms, respectively. The notations $<ij>$, $<ijk>$, etc., in Eq.~(\ref{model}) indicate that the summation is restricted to tuples of nearest-neighbor spins for HOI, which is justified by the fourth-order perturbative origin of these interactions. Dipole–dipole interactions are not included in our model, as their contribution is expected to be negligible in the monolayer Janus systems considered here. To construct a realistic spin model, all the magnetic interaction parameters, including HOI, are obtained from DFT calculations (see Table \ref{magnetic_parameters}). Throughout this work, all atomistic spin simulations are performed in $120 \times 120$ lattices with periodic boundary conditions using the \textsc{spinaker} code.

\subsection*{B. Minimum energy path calculations}

Minimum energy paths (MEPs) for the annihilation of magnetic solitons into the field-polarized state, hereafter referred to as the ferromagnetic (FM) state, are calculated using the geodesic nudged elastic band (GNEB) method~\cite{bessarab2015method}. The GNEB method determines the MEP in the space of magnetic configurations connecting a metastable initial state (ini) to a final state (fin). The first-order SP along the MEP is refined using the climbing-image procedure~\cite{bessarab2015method}. In this work, we consider two energy barriers associated with the collapse ($\Delta E$) and nucleation ($\Delta E^{\text{fin}\rightarrow\text{ini}}$) of magnetic solitons. These two barriers govern the stability of magnetic solitons and are therefore key quantities for spintronics applications. The corresponding collapse energy barrier for the transition from the initial soliton state to the FM state is
\begin{equation}
\Delta E
= E_{\text{SP}}-E_{\text{ini}}~,
\end{equation}
while the nucleation energy barrier for the reverse transition from the final state (FM state) to the initial state (soliton state) is
\begin{equation}
\Delta E^{\text{fin}\rightarrow\text{ini}}
= E_{\text{SP}}-E_{\text{fin}}~.
\end{equation}
Here, $E_{\text{ini}}$, $E_{\text{fin}}$, and $E_{\text{SP}}$ denote the energies of the initial state, final state, and SP, respectively.

\section{RESULTS AND DISCUSSION}

As a representative platform, we consider the 2D vdW Janus monolayer MnPCl with a triangular lattice (see Supplementary Note 1 in Supplementary Material (SM) \cite{supplmat} for the atomic structure). Recently, Janus monolayers have been extensively studied theoretically~\cite{Hongxin2020,changsong2020,megha2025} because the monolayer limit, combined with intrinsic broken inversion symmetry, provides a simple setting for a new skyrmion platform. Importantly, recent breakthroughs in the synthesis of magnetic Janus monolayers pave the way for the experimental realization and investigation of such phenomena in this emerging class of materials~\cite{xu2025unusual,nie2024regulated}. Our calculations are done using a state-of-the-art multiscale approach that combines DFT calculations with atomistic spin simulations (see ``METHODS"). Because the magnetic properties of Janus monolayers have already been extensively investigated by DFT, we focus here on atomistic spin simulations. The magnetic interaction parameters, extracted from~Ref. \cite{Hongxin2023}, are adapted to our spin model and summarized in Table~\ref{magnetic_parameters}. Details of the Hamiltonian mapping are provided in Supplementary Note~1 of SM \cite{supplmat}. The spin spiral dispersion obtained from the DFT magnetic interactions is extremely flat near the $\bar{\Gamma}$ point (Supplementary Note 1 of SM \cite{supplmat}), which indicates the possibility of stabilizing sub-10 nm solitons without the presence of external magnetic fields.

\begin{figure*}[tb]
	\centering
	\includegraphics[width=1.0\linewidth]{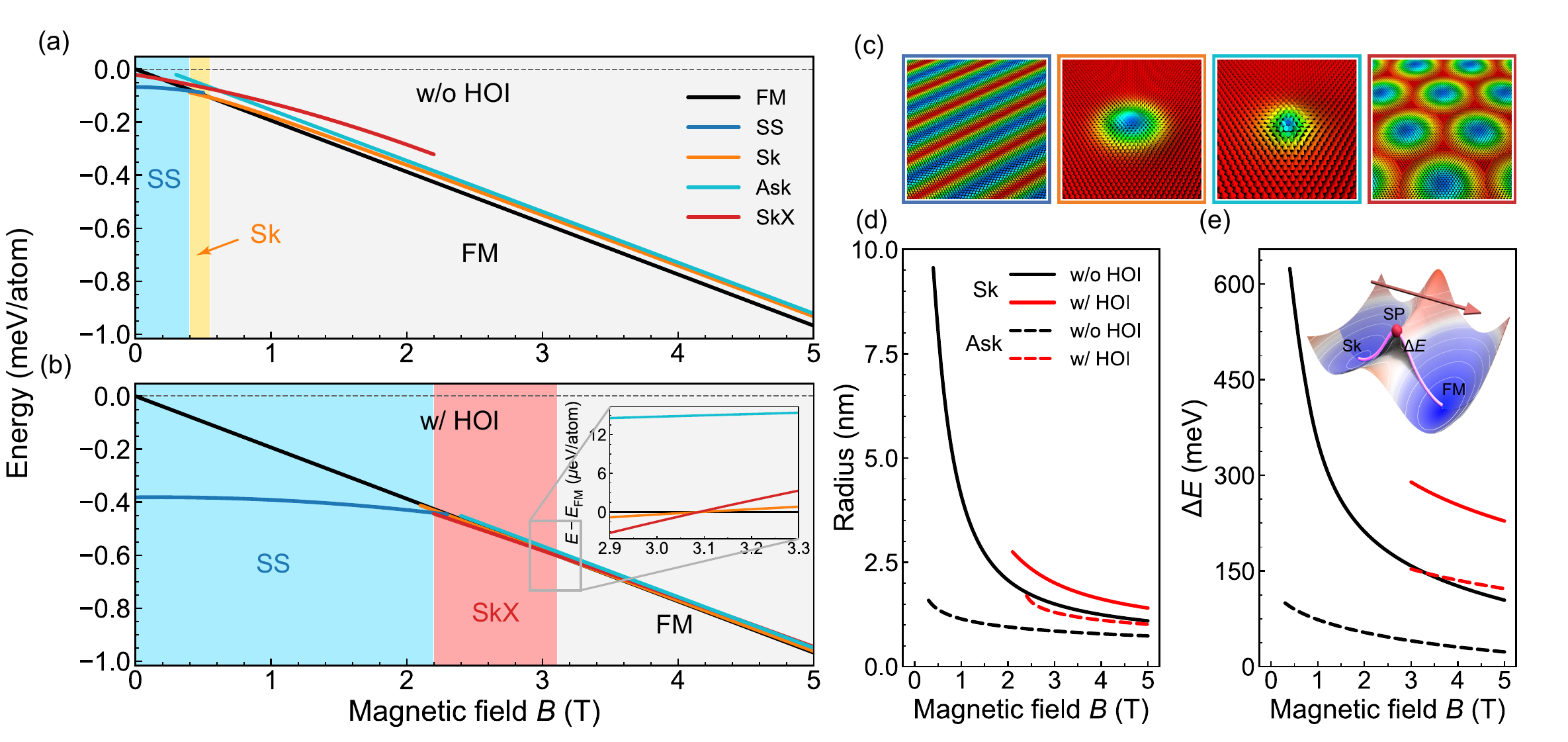}
	\caption{
    Field-dependent phase stability and collapse barriers of topological spin textures without and with HOI.
    (a-b)~Energies of the ferromagnetic (FM), spin spiral (SS), skyrmion (Sk), antiskyrmion (Ask), and skyrmion lattice (SkX) states as functions of $B$, obtained without and with HOI. Sk denotes a single skyrmion per simulation cell, that is, the most dilute lattice periodicity considered here, whereas SkX refers to a dense skyrmion lattice. Energies are given relative to the FM state at $B=0$, and shaded regions indicate the ground state in each field range. The inset in panel (b) highlights the energy difference $E-E_{\text{FM}}$ near the SkX–FM phase boundary.
    (c)~Spin configurations used as initial states: SS, Sk, Ask and SkX, with frame colors as in the legend of panel (a).
    (d)~Magnetic-field dependence of the radius of isolated skyrmions (solid lines) and antiskyrmions (dashed lines) without (black) and with HOI (red).
    (e)~Corresponding energy barriers $\Delta E$ against collapse into the FM background, without (black) and with HOI (red). 
    Inset: Schematic energy landscape defining $\Delta E$ as the energy difference between the saddle point (SP) and the Sk state. All simulations are performed on a $120\times120$ spin lattice with periodic boundary conditions.
    }
    \label{phase}
\end{figure*}

\subsection*{A. Elementary and high-$Q$ spin textures stabilized by HOI}
Using the soliton model described in ``METHODS", we initialized isolated skyrmions ($Q=-1$), antiskyrmions ($Q=1$), and skyrmioniums ($Q=0$) in the field-polarized background and relaxed them with the full set of DFT-derived magnetic interactions, which contains the fourth-order terms shown in Fig.~\ref{overview}(b). All three remain metastable, so MnPCl supports the complete set of elementary textures, namely low-$Q$ states ($|Q|\le1$), with diameters below 10 nm [Fig.~\ref{overview}(a)]. We further obtain a much richer family of solitons with high topological charge [Fig.~\ref{overview}(c)]. These include skyrmion and antiskyrmion rings as well as skyrmion bags, in which several inner skyrmions share a common outer skyrmion, which together obey charges up to $|Q| =5$. All low-$Q$ textures in Fig.~\ref{overview} are relaxed with the same interaction parameters, including HOI, summarized in Table~\ref{magnetic_parameters}. In the following, to investigate the role of HOI, we consider two parameter sets, $\mathcal{O}_{\text{w/o}}= \{\{J_{ij}\},\{\mathbf{D}_{ij}\}, K_u\}$ without HOI and $\mathcal{O}_{\text{w/}}= \{\{J_{ij}\},\{\mathbf{D}_{ij}\}, K_u, B_1, Y_1, K_1\}$ with HOI. This comparison allows us to directly address the role of HOI in stabilizing the spin textures.

\subsection*{B. Phase diagram and stability of isolated skyrmions and antiskyrmions}

\begin{figure*}[tbp]
	\centering
	\includegraphics[width=1.0 \linewidth]{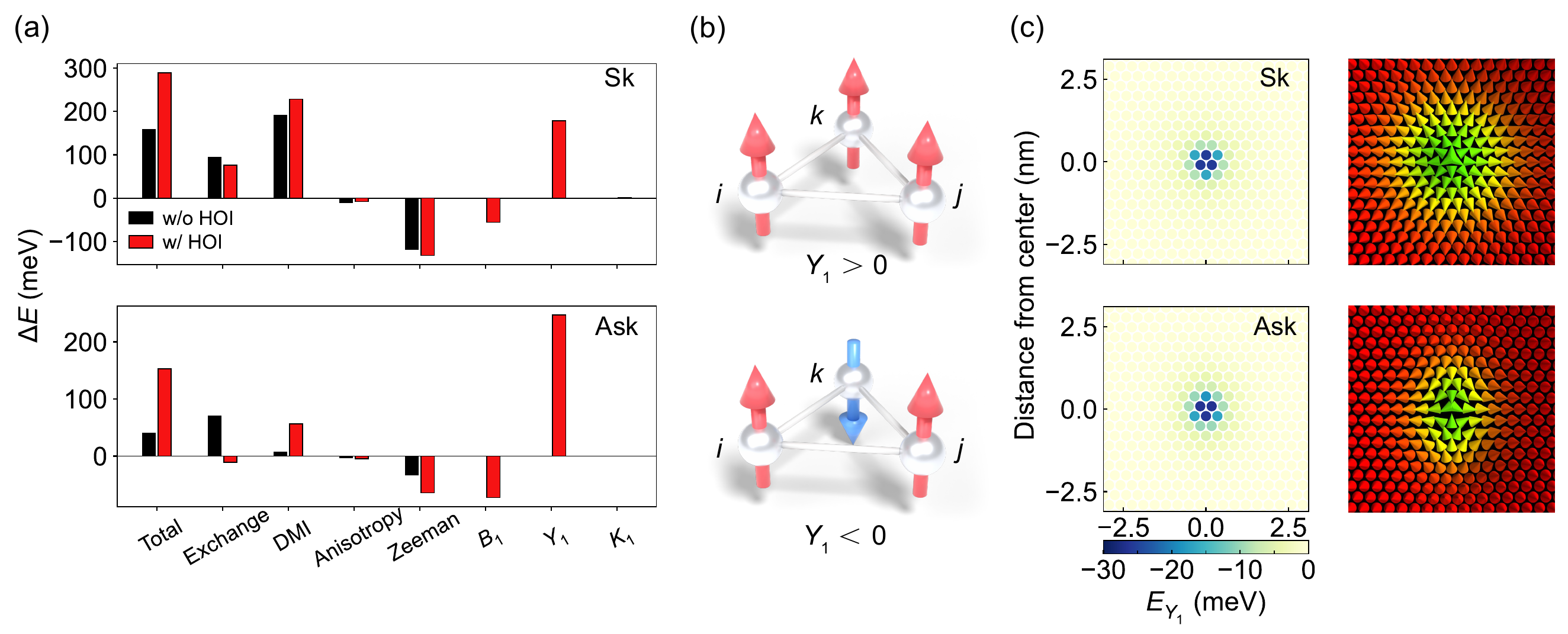}
	\caption{
    Energy barrier decomposition and role of the four-spin three-site interaction $Y_1$ in soliton stability.
    (a)~Decomposition of the collapse barrier of Fig.\ref{phase} at $B=3$ T into individual energy terms, $\Delta E=E_\textrm{SP}-E_\textrm{ini}$, for the skyrmion (Sk, top) and the antiskyrmion (Ask, bottom), calculated without (black) and with (red) HOI. $B_1$, $Y_1$, and $K_1$ denote the biquadratic, four-spin three-site, and four-spin four-site terms of Table~\ref{magnetic_parameters}.
    (b)~Schematic illustration of the energetically favored local spin arrangements on a triangular plaquette for positive and negative $Y_1$. A positive $Y_1$ favors the same sign of the scalar products of the $(i,j)$ and $(j,k)$ spin pairs, whereas a negative $Y_1$ favors opposite signs and thus acts as an 
    effective exchange frustration.
    (c)~Atom-resolved $Y_1$ energy at the SP for the skyrmion (top) and the antiskyrmion (bottom). The left panels show the real-space $Y_1$ energy density, $E_{Y_1}$, relative to the FM state, while the right panels show the corresponding spin configurations at the SP. The $Y_1$ contribution is strongly localized around the soliton core. 
    }
    \label{decomposition} 
\end{figure*}

We first investigate the phase diagram of the elementary $|Q|=1$ solitons, both without and with HOI, to determine the magnetic ground state and identify the field range in which isolated skyrmions and antiskyrmions are stable. Comparing the energies of the relaxed minimum spin spirals (SS), skyrmions (Sk), antiskyrmions (Ask), skyrmion lattices (SkX), and ferromagnetic (FM) states yields the magnetic phase diagrams shown in Fig.~\ref{phase}(a-b). Representative spin textures of the competing magnetic states are shown in Fig.~\ref{phase}(c). 
Without HOI, increasing $B$ drives the ground state from the SS phase to the Sk phase and ultimately to the field-polarized FM state [Fig.~\ref{phase}(a)]. Incorporating HOI qualitatively changes the phase diagram [Fig.~\ref{phase}(b)]. The SS phase extends over a substantially broader field range, while the Sk phase is replaced by a denser SkX phase (the relative stability of SkX states with different lattice periodicities is discussed in Supplementary Note 3 of SM \cite{supplmat}). Both changes originate from the negative $Y_1$, which lowers the energy of noncollinear spins [Fig.~\ref{decomposition}(b)]. This gain scales with the noncollinear area, so it favors the spin spiral and the denser SkX. As $B$ increases, the energy difference between the SkX and FM phases continuously decreases and vanishes at the critical field of $B_{\text{c}} \approx 3$~T, above which the FM becomes the ground state. Near this transition, the isolated skyrmion lies within a few $\mu$eV per atom of the SkX and FM states [inset of Fig.~\ref{phase}(b)]. For $B>B_{\text{c}}$, isolated skyrmions and antiskyrmions exist as metastable excitations within the FM background.

Notably, both skyrmions and antiskyrmions remain compact throughout the investigated field range, with radii of only about 1.8~nm and 1.5~nm, respectively, near $B=3$~T [Fig.~\ref{phase}(d)]. Their field dependence, however, differs qualitatively. The skyrmion contracts rapidly with increasing $B$, whereas the antiskyrmion radius remains nearly $B$-independent. This difference originates from the distinct interactions governing their equilibrium sizes. The skyrmion radius results from a competition between the DMI and the Zeeman energy. The former lowers the energy in proportion to the length of the circular wall bounding the skyrmion, whereas the latter penalizes the reversed core. As $B$ increases, the growing Zeeman cost progressively suppresses the DMI-driven expansion and the skyrmion shrinks~\cite{Romming2015}. By contrast, the antiskyrmion size is determined mainly by exchange frustration~\cite{Malottki2017} and is therefore only weakly affected by $B$. Moreover, the alternating wall chirality causes favorable and unfavorable DMI contributions to largely compensate, leaving only a minor net DMI stabilization~\cite{Hoffmann2017}. 
Accordingly, the two radii become comparable at the upper end of the field range studied here [Fig.~\ref{phase}(d)], even though they are set by different interactions.

To quantify the stability of solitons, the collapse energy barrier $\Delta E$, preventing skyrmions and antiskyrmions from collapsing into the FM state along the MEP, is obtained using the GNEB method \cite{bessarab2015method} (see ``METHODS" for computational details). As isolated skyrmions and antiskyrmions become metastable only for $B\geq3$~T when HOI are included, we evaluate $\Delta E$ in this field range. Without HOI, the skyrmion barrier decreases from 624~meV at 0.4~T to 104~meV at 5~T, whereas the antiskyrmion barrier remains below 100~meV throughout the investigated field range [Fig.~\ref{phase}(e)]. Even without HOI, such collapse barriers exceed those reported for other 2D vdW magnets by one to two orders of magnitude~\cite{Dongzhe2022_fgt,Samir_2023,shiwei2026,zhang2026electric} and are already comparable to those of state-of-the-art ultrathin films~\cite{paul2020role,goerzen2023lifetime}. By contrast, HOI enhance $\Delta E$ substantially while leaving the soliton size nearly unchanged. At $B = 5$~T, the skyrmion barrier increases from 104 to 228~meV upon inclusion of HOI, while the antiskyrmion barrier increases from 23 to 122~meV [Fig.~\ref{phase}(e)]. These increases correspond to approximately twofold and fivefold enhancements for skyrmions and antiskyrmions, respectively, relative to the cases without HOI. We next identify the microscopic origin of this nontrivial behavior.

\subsection*{C. Origin of the HOI-induced stabilization}

To elucidate the microscopic origin of the large $\Delta E$ and its pronounced enhancement by HOI, we decompose the collapse barrier into individual interaction terms [Fig.~\ref{decomposition}(a)]. Without HOI (black bars), the stability of skyrmions originates mainly from the combined effects of DMI and exchange frustration, whereas the Zeeman interaction favors the FM state. Antiskyrmions, in contrast, gain almost no stabilization from the DMI due to symmetry constraints. In the continuum limit, the interfacial DMI energy of a radially symmetric soliton can be written as
\begin{equation}\label{dmi}
   E_{\text{DMI}}=2 \pi \mathcal{D} \delta_{\nu, 1} \cos \gamma \int_{0}^{\infty}\left(\frac{\partial \Theta}{\partial \rho}+\frac{1}{\rho} \sin \Theta \cos \Theta\right) \rho \mathrm{d} \rho ~,
\end{equation}
where $\Theta(\rho)$ defines the radial profile of solitons, $\gamma$ is the helicity, $\mathcal{D}$ is the microscopic DMI, and $\nu$ is the vorticity (see ``METHODS" for details). The factor $\delta_{\nu,1}$ shows that the DMI contributes only to textures with $\nu=1$, and therefore vanishes for antiskyrmions with $\nu=-1$~\cite{Koshibae2016,Hoffmann2017}.

The inclusion of HOI qualitatively changes this energy balance (red bars). The four-spin three-site interaction $Y_1$ contributes the most to $\Delta E$, while the biquadratic term $B_1$ slightly favors the FM state. Since $Y_1$ is insensitive to chirality, it provides a comparable enhancement of $\Delta E$ of about 200 meV for both skyrmions and antiskyrmions. The effect of $Y_1$ is not limited to an additional energy contribution. Instead, it fundamentally reshapes the energy landscape and the corresponding collapse pathway. The exchange contribution is strongly reduced and the DMI contribution enhanced, not only at the SP but along the entire MEP (Supplementary Note~2 in SM \cite{supplmat}).

\begin{figure*}[tbp]
	\centering
	\includegraphics[width=1.0\linewidth]{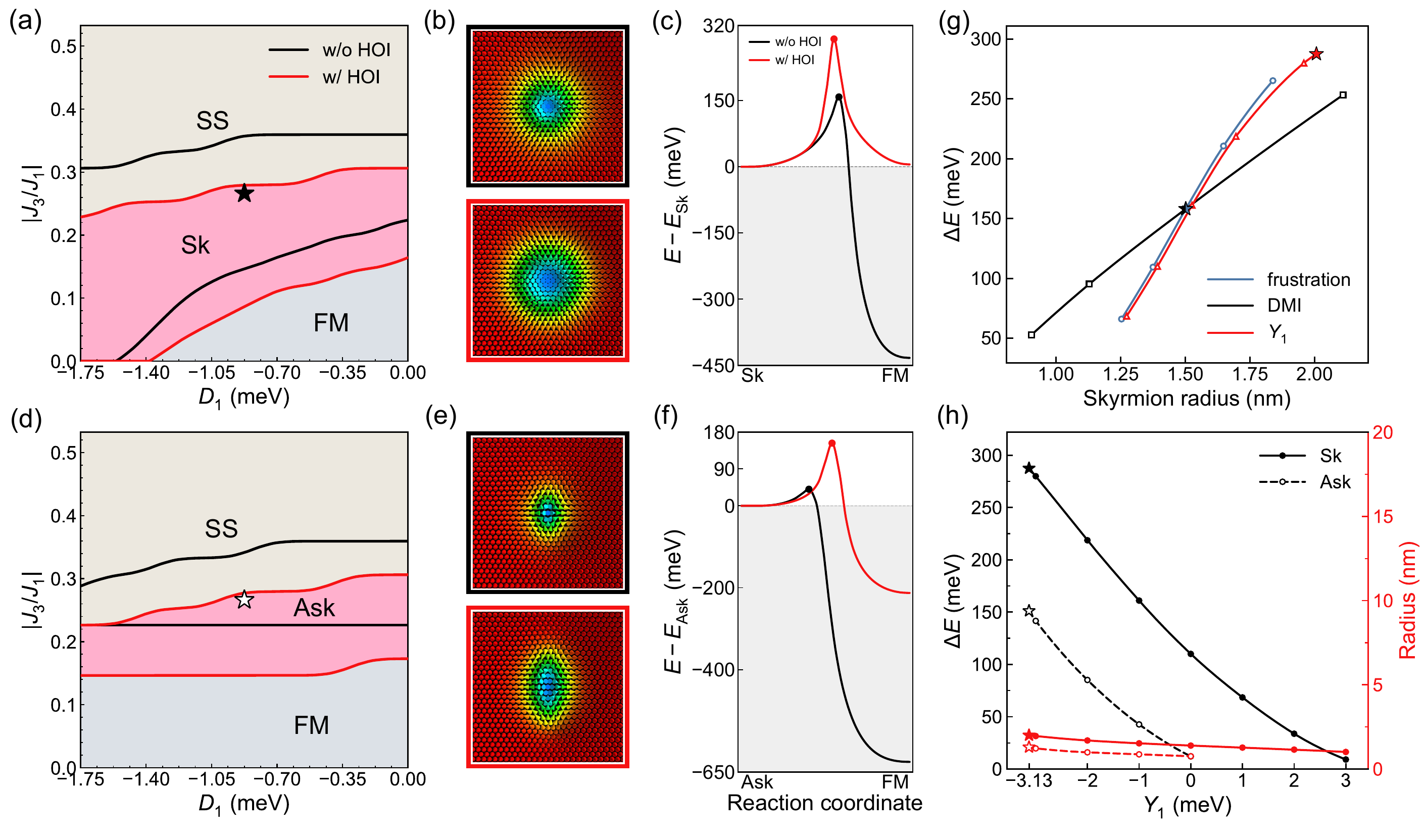}
	\caption{
    Tuning skyrmion and antiskyrmion stability through exchange frustration, DMI, and HOI.
    (a)~Phase diagrams at $B=3$~T as functions of the nearest-neighbor DMI $D_1$ and exchange frustration $|J_3/J_1|$, calculated without (black boundaries) and with (red boundaries) HOI, for skyrmions. The stars denote the full DFT parameter set used throughout this work.
    (b)~Equilibrium skyrmion spin textures at the parameter set marked by the star, calculated without (top) and with (bottom) HOI.
    (c)~Corresponding MEPs for the skyrmion collapse, without (black) and with (red) HOI, filled circles mark the SP. HOI significantly enhance the collapse barrier.
    (d-f)~Same as (a-c), but for antiskyrmions.
    (g)~Energy barrier $\Delta E$ of the isolated skyrmion versus its radius, obtained by individually varying the frustration ratio (blue), DMI (black), and $Y_1$ (red). The highlighted point on each curve marks the DFT value of that parameter, so the blue and black ones coincide at the reference without HOI, while the red one is the full DFT parameter set including HOI.
(h)~$Y_1$-dependence of $\Delta E$ (left axis, black) and radius (right axis, red) for Sk (solid) and Ask (dashed). Black and red distinguish the two axes. Filled and open circles indicate Sk and Ask, respectively. Stars mark the DFT parameter, $Y_1=-3.13$~meV.
    }
    \label{tuning}
\end{figure*}

This behavior can be traced to the frustration induced by the $Y_1$ term, as seen directly from its energy expression $H_{3s}=-2Y_1\sum(\mathbf{m}_i\!\cdot\!\mathbf{m}_j)(\mathbf{m}_j\!\cdot\!\mathbf{m}_k)$. As illustrated in Fig.~\ref{decomposition}(b), the energy is determined by the product of the scalar products of the $(i,j)$ and $(j,k)$ spin pairs. For $Y_1>0$, the energy is minimized when the two scalar products have the same sign, favoring collinear spin alignments. In contrast, for $Y_1<0$, the energy is minimized when they have opposite signs [lower panel of Fig.~\ref{decomposition}(b)]. 
Extending this preference throughout the triangular lattice introduces effective frustration, favoring noncollinear magnetic textures. Since $Y_1$ becomes the dominant contribution along the MEP, the spin structure reconstructs itself to lower the $Y_1$ energy. Because this reconstruction favors noncollinear configurations, it indirectly reduces the exchange contribution to $\Delta E$, while the modified spin profile increases the DMI contribution. This effect is particularly pronounced for antiskyrmions, where the redistribution along the collapse pathway even reverses the sign of the exchange contribution.

This frustration acts locally at the SP [Fig.~\ref{decomposition}(c)]. The atom-resolved $Y_1$ energy is concentrated on a small group of spins around the collapsing core for both skyrmions and antiskyrmions. The spins around the core undergo the largest rotations along the collapse pathway (Fig.~\ref{decomposition}(c), right), causing the negative $Y_1$ interaction to raise the energy of the SP most efficiently. The enhanced collapse barrier, therefore, originates from a localized increase in energy at the SP rather than from a global modification of the energy landscape. Both solitons undergo the same radial collapse without and with HOI, so HOI increase the energy barrier significantly without altering the collapse mechanism.

The phase diagrams in the $D_1$--$|J_3/J_1|$ plane show that HOI shift the stability windows of both skyrmions and antiskyrmions towards weaker bilinear exchange frustration [Fig.~\ref{tuning}(a),(d)]. HOI, therefore, provide an effective source of frustration that complements the competing bilinear exchange interactions. At the DFT parameters of MnPCl (stars), both the skyrmion and antiskyrmion are stable. More generally, placing the DFT-derived magnetic interaction parameters of the Mn- and Cr-based Janus families reported in Ref.~\cite{Hongxin2023} on these phase diagrams reveals several additional soliton-host materials. CrSeCl, MnPBr, and CrSeBr are predicted to host both skyrmion and antiskyrmion states, whereas CrTeCl, CrTeBr, and MnAsBr are predicted to reside in the SS regime, from which isolated skyrmions and antiskyrmions are expected to emerge under external magnetic fields.

For MnPCl, the models with and without HOI support solitons of similar size and texture [Fig.~\ref{tuning}(b),(e)]. The antiskyrmion, however, is elongated in contrast to the nearly circular skyrmion, and the inclusion of HOI substantially enhances this elongation. Energy density analysis (Supplementary Note~6 in SM \cite{supplmat}) traces this behavior to the interplay between the DMI and the four-spin three-site interaction $Y_1$. The DMI is the only interaction that generates an anisotropic energy distribution along the circular antiskyrmion domain wall, whereas $Y_1$ is almost isotropic. Because the $Y_1$ and bilinear exchange contributions are comparable in form and opposite in sign, a growing $|Y_1|$ cancels part of the isotropic exchange and lets the weak DMI anisotropy set the shape. At the DFT parameters, the ratio of the soliton radii along $y$ and $x$ reaches $\varepsilon = R_y/R_x \approx 1.54$ (Supplementary Note~6 in SM \cite{supplmat}). In contrast, the collapse energetics of the two models differ strongly. The MEPs computed with HOI reach a much higher saddle point for both solitons [Fig.~\ref{tuning}(c),(f)].

\begin{figure*}[tp]
	\centering
	\includegraphics[width=1.0\linewidth]{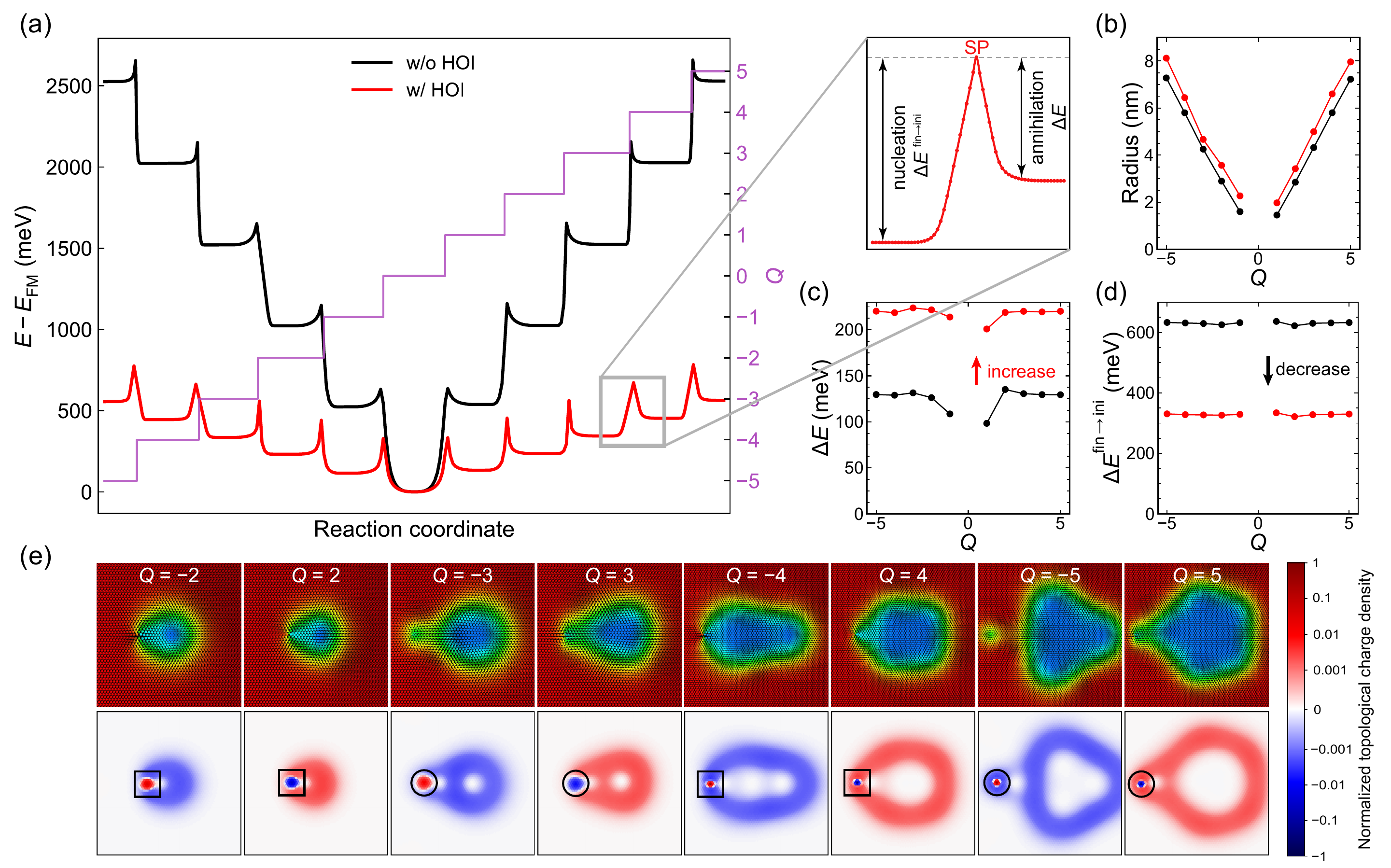}
	\caption{
    MEPs and collapse mechanisms of high-$Q$ skyrmion and antiskyrmion rings at zero field.
    (a)~Total energy relative to the FM state along MEPs connecting the ring states from $Q=-5$ to $Q=5$, calculated without (black) and with (red) HOI at zero magnetic field. The violet staircase shows the evolution of $Q$ along the reaction coordinate (right axis). Inset: definition of the energy barrier $\Delta E$ and the nucleation barrier $\Delta E^{\text{fin}\rightarrow\text{ini}}$ (SP relative to the collapsed state) at a single transition.
    (b)~Radius of the ring states versus $Q$, without and with HOI, all textures remain well below 20~nm in diameter.
    (c-d)~$Q$ dependence of $\Delta E$ and $\Delta E^{\text{fin}\rightarrow\text{ini}}$, without and with HOI.  
    (e)~Spin configurations and normalized topological charge density at the BP for the ring states with $|Q|=2$--$5$, obtained with HOI. For even $Q$, the collapse proceeds through a chimera mechanism characterized by a local reversal of the topological charge density (squares). For odd $Q$, the transition occurs via a previously unreported \textit{ejection} collapse mechanism, in which a localized soliton detaches from the ring (circles).
    All calculations are performed with the DMI reduced to 5\% of its DFT value.}
    \label{rings} 
\end{figure*}

Plotting $\Delta E$ against the skyrmion radius separates the routes by which the DMI, the bilinear exchange frustration $|J_3/J_1|$, and $Y_1$ enhance stability [Fig.~\ref{tuning}(g)]. Independent variations of $Y_1$ and $|J_3/J_1|$ produce nearly parallel curves with steep slopes, whereas the DMI curve has a much smaller slope, so any barrier gained through the DMI comes with a much larger increase in the skyrmion radius. The $Y_1$ curve closely follows the $|J_3/J_1|$ curve, consistent with the effective frustration introduced by the $Y_1$ term [Fig.~\ref{decomposition}(b)]. Instead of enlarging the soliton, exchange frustration acts directly on the collapse pathway by raising the energy cost of the large-angle spin rotations localized at the SP~\cite{Malottki2017}. Exchange frustration and HOI therefore provide access to a higher collapse barrier than DMI, with little or no accompanying increase in soliton size.

As $Y_1$ becomes more negative, $\Delta E$ rises from 9 to 288~meV for the skyrmion and from 12 to 151~meV for the antiskyrmion, while both radii remain between 1 and 2~nm [Fig.~\ref{tuning}(h)]. 
The antiskyrmion reaches the lower end of its range as $Y_1$ approaches zero and is no longer metastable beyond that point, whereas skyrmions persist up to $Y_1=3$~meV, although with a substantially reduced $\Delta E$. The antiskyrmion, unlike the skyrmion, cannot be stabilized by the DMI [cf. Eq.~(\ref{dmi})], even by gaining energy from elongation \cite{kuchkin2020turning}. It therefore relies on the additional frustration supplied by $Y_1$. Calculations at more negative $Y_1$, beyond the range shown in Fig.~\ref{tuning}(h), give a lower $\Delta E$. The DFT value of $Y_1=-3.13$ meV [red star in Fig.~\ref{tuning}(g)] therefore places MnPCl near the maximum stability of both textures. 

\subsection*{D. MEPs of high-$Q$ skyrmion and antiskyrmion rings}

Isolated high-$Q$ textures are challenging to stabilize because their circumference necessarily contains segments in which the spins rotate in opposite senses on going outward, which a homogeneous DMI cannot favor simultaneously. Existing strategies therefore realize them through dipolar interactions in multilayers~\cite{hassan2024dipolar}, spatially engineered DMI landscapes~\cite{Niu2025}, confinement by dipolar fields and artificial anisotropy defects~\cite{Kern2025}, or as skyrmion bundles in bulk chiral magnets~\cite{zhang2024stable}. These approaches, however, produce high-$Q$ textures that are typically hundreds of nanometers in size, which limits their applicability in compact spintronic devices. By contrast, our strategy exploits the frustration supplied by $Y_1$, which is intrinsic and insensitive to the sense of spin rotation. To suppress the residual chiral preference of the DMI, we reduce it to 5\% of its DFT value while leaving all other interactions unchanged (Supplementary Note~4 in SM \cite{supplmat}). This value is a compromise. Without DMI, the rings remain metastable, but their collapse barriers decrease by about one order of magnitude, whereas at 10\% of the DFT value, the $Q=-2$ and $Q=-5$ textures no longer relax into ring states once HOI are included (Supplementary Note~5 in SM \cite{supplmat}). Such a reduction may also be experimentally accessible, since the DMI is known to decrease with temperature more rapidly than the exchange interactions \cite{kim2018correlation,Mankovsky2020}.

In this setup, zero-field skyrmion and antiskyrmion rings with arbitrary $Q$ can be stabilized. Fig.~\ref{rings}(a) illustrates the corresponding MEPs for rings with $Q=-5$ to $Q=5$, 
where $Q$ changes by one at each transition (violet staircase). Without HOI, this sequence spans nearly 2.5 eV, placing the $Q=\pm5$ rings far above the FM state. Including HOI compresses the entire energy spectrum to approximately 0.55 eV. The collapse energy barrier $\Delta E$ protects a ring against the loss of one unit of topological charge, whereas the nucleation energy barrier $\Delta E^{\text{fin}\rightarrow\text{ini}}$ must be overcome to reform the ring from the collapsed state (see ``METHODS"). A $Q=-5$ ring first transforms into a $Q=-4$ ring rather than collapsing directly into the FM state. Complete annihilation, therefore, proceeds through five successive thermally activated transitions.

\begin{figure*}[tp]
	\centering
	\includegraphics[width=1.0\linewidth]{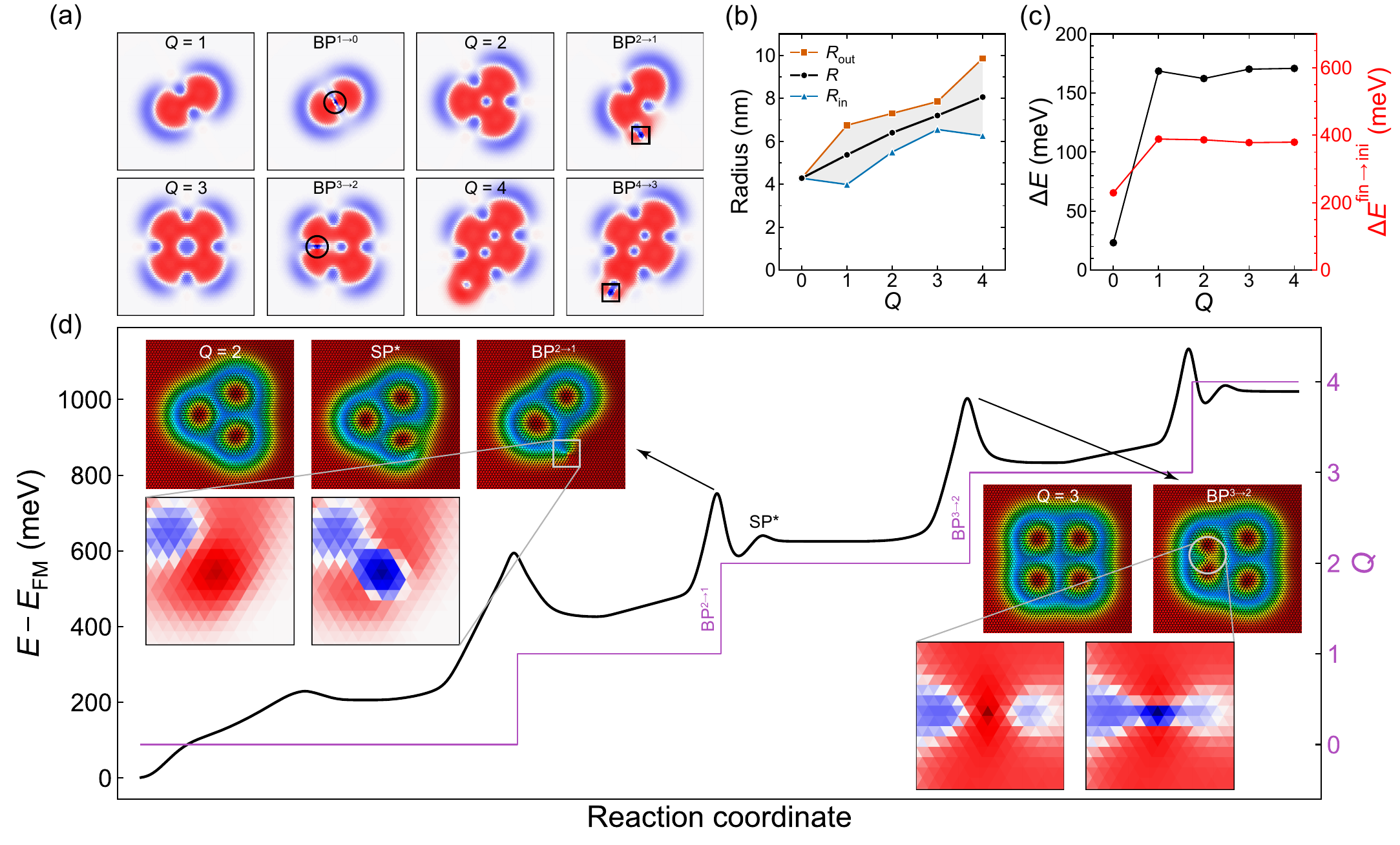}
	\caption{
    MEPs and collapse mechanisms of skyrmion bags at $B=3$ T.
    (a)~Normalized topological charge density of skyrmion bags with $Q=1$--$4$ (left of each pair) and at the corresponding $\mathrm{BP}^{Q\rightarrow Q-1}$ (right of each pair). Two distinct collapse mechanisms are identified. For odd $Q$, two inner skyrmions merge, and the topological charge density reverses at the merging point inside the bag (circles). For even $Q$, an inner skyrmion couples to the outer boundary and triggers a local reversal at a corner kink of the outer wall (squares). We term these two previously unreported collapse mechanisms \textit{combining} collapse and \textit{kink} collapse, respectively.
    (b)~Inner radius $R_{\text{in}}$, outer radius $R_{\text{out}}$ and effective radius $R=(R_{\text{in}}+R_{\text{out}})/2$ of skyrmion bags as a function of $Q$, where $R_{\text{in}}$ and $R_{\text{out}}$ are the inner and outer $m_z=0$ contours of the texture.
    (c)~Annihilation energy barrier $\Delta E$ (black, left axis) and nucleation barrier $\Delta E^{\text{fin}\rightarrow\text{ini}}$ (red, right axis) of skyrmion bags for $Q=0$--$4$, where $Q=0$ corresponds to the skyrmionium state.
    (d)~Total energy relative to the FM state along MEPs connecting the bag states up to $Q=4$, the violet staircase shows $Q$ (right axis). Spin configurations illustrate the two mechanisms. The $Q=2\rightarrow1$ transition (top left), following the \textit{kink} mechanism, proceeds through an intermediate local maximum (SP$^{*}$), at which an inner skyrmion first touches the outer boundary, before crossing the true SP towards $\text{BP}^{2\rightarrow1}$. For the $Q=3\rightarrow2$ transition (bottom right), the collapse follows the \textit{combining} mechanism ($Q=3$ and $\text{BP}^{3\rightarrow2}$). Insets show the local charge density immediately before and after the charge reversal at the kink (left) and at the merging point (right). We emphasize that skyrmion bags are stabilized only in the presence of HOI.
    }
    \label{bags}
\end{figure*}

The ring radius increases nearly linearly with $|Q|$, from 1.5 nm for $|Q|=1$ to 8.1 nm for $|Q|=5$, and is almost unchanged by HOI [Fig.~\ref{rings}(b)]. Interestingly, the inclusion of HOI enhances $\Delta E$ while substantially reducing $\Delta E^{\text{fin}\to\text{ini}}$ for every ring state [Fig.~\ref{rings}(c-d)]. Relative to the collapsed state, both the ring and the SP contain an additional segment of the ring wall, whose noncollinearity is favored by the negative $Y_1$. The ring gains the full energetic stabilization because its wall remains close to the optimal canting angle. At the SP, however, part of this wall segment is forced into the large-angle rotations penalized by $Y_1$, as discussed before. HOI therefore lower the energies of both the ring and the SP, but stabilize the ring more strongly. 

Both barriers are nearly independent of $Q$ across the entire ladder [Fig.~\ref{rings})c-d], despite the rapid increase in both the energy and the radius of the rings with $|Q|$. This indicates that the collapse is mainly governed by a local rather than a collective instability. The topological charge density (see ``METHODS") at the image containing a Bloch-point-like singularity (BP) confirms this picture, showing that the topological rearrangement is confined to the edge of the ring [Fig.~\ref{rings}(e)]. 
A high-$Q$ ring is thus nearly as stable as an isolated $|Q|=1$ soliton. The collapse barriers range from 98 to 224 meV [Fig.~\ref{rings}(c)], one to two orders of magnitude larger than those recently reported in a vdW heterostructure~\cite{shiwei2026}.

The collapse mechanism of high-$Q$ rings depends on the parity of $Q$. Two collapse mechanisms are known for low-$Q$ solitons. One is the radial collapse, in which the soliton shrinks homogeneously until its central spins rotate into the film plane~\cite{bessarab2015method,Rohart2016}. The other is the chimera collapse, in which the soliton undergoes little overall contraction and the topological charge density instead reverses locally at a BP displaced from its center~\cite{Meyer2019,Desplat2019}. Both mechanisms have been observed experimentally in ultrathin films~\cite{Muckel2021}. 
The rings never shrink, which rules out the radial route for all $|Q| \ge 2$. For even $Q$ the transition is of the chimera type. The reversed core forms within the ring wall and remains embedded in it, and the ring keeps its connectivity throughout [squares in Fig.~\ref{rings}(e)]. A similar boundary collapse was recently found for high-$Q$ bimeron rings \cite{shiwei2026}, and HOI already select the chimera pathway at $|Q|=1$ \cite{paul2020role}. For odd $Q$, the BP configurations show a different topology. The reversed core is separated from the ring by a belt of vanishing topological charge density [white regions in Fig.~\ref{rings}(e)]. The reversal occurs inside a localized soliton that has detached from the ring and carries one unit of topological charge away [circles in Fig.~\ref{rings}(e)]. We term this previously unreported mechanism \textit{ejection} collapse. The configurations in Fig.~\ref{rings}(e) are taken from the model with HOI, whose inclusion does not alter the collapse mechanisms or the parity rule governing them.

Because the topological charge changes by one at each transition, this parity rule determines the entire decay sequence. For instance, a $Q=\pm5$ ring undergoes alternating \textit{ejection} and chimera collapse events before the final $|Q|=1$ soliton annihilates through radial collapse. A single decay sequence, therefore, samples all three mechanisms. Our predictions can be tested experimentally using either SP-STM \cite{Muckel2021} or magnetic force microscopy \cite{koraltan2025signatures}.

\subsection*{E. MEPs of high-$Q$ skyrmion bags}

Skyrmion bags reach an arbitrary $Q$ with a different architecture. A skyrmion shell encloses $Q+1$ inner skyrmions, so every wall winds once and with the same rotational sense \cite{Foster2019, Rybakov2019, Kuchkin2020}. Unlike skyrmion rings, they therefore satisfy the chiral constraint imposed by the DMI. Skyrmion bags have been observed in chiral magnets \cite{Tang2021, yang2024embedded,zhang2024stable}, vdW ferromagnets \cite{Powalla2023, Hu2026}, and moiré superlattices \cite{schwab2025skyrmion}, with sizes ranging from tens to hundreds of nanometers. However, despite recent progress in stabilizing and engineering skyrmion bags~\cite{Kuchkin2020,Foster2019,liu2025room,Kern2025,wu2026current}, the collapse barriers that determine their thermal stability, which is a key requirement for applications, remain unknown. In our work, we present the first calculations of the energy barriers of high-$Q$ skyrmion bags with the full set of DFT parameters at $B=3$ T. Without HOI, no skyrmion bags survive after energy minimization. With HOI, the complete family of skyrmion bags, from the skyrmionium ($Q=0$), the smallest bag, to arbitrary $Q \geq 1$, is stabilized. Since a bag with charge $Q$ encloses $Q+1$ inner skyrmions, the skyrmionium is its smallest member. A bag with $Q<0$ would contain no inner skyrmion and reduces to the isolated skyrmion. Fig.~\ref{bags}(a) shows the topological charge density of each relaxed bag up to $Q=4$, together with that at the BP driving the corresponding $Q\rightarrow Q-1$ transition, denoted $\mathrm{BP}^{Q\rightarrow Q-1}$. 

The effective radius $R=(R_{\text{in}}+R_{\text{out}})/2$ of the inner and outer $m_z = 0$ contours grows from 4.3 nm for the skyrmionium to 8.1 nm at $Q=4$, while the outer radius remains below 10 nm [Fig.~\ref{bags}(b)]. As a result, all skyrmion bags remain below 20 nm in diameter, far smaller than previously reported skyrmion bags, whose characteristic sizes range from several tens of nanometers to the micrometer scale~\cite{Rybakov2019,Jiang2025,liu2025room,wu2026current}.
Such compact-size solitons originate from the extremely flat spin-spiral dispersion near the FM state, as shown in Supplementary Note 1 of SM \cite{supplmat}. Their thermal stability shows one exception. The skyrmionium ($Q=0$) decays directly into the FM state with a barrier of only 23 meV (see Fig.~\ref{bags}(c), Supplementary Note~7 and Animation 1 in SM \cite{supplmat}). All bags with $Q\ge1$ instead exhibit nearly constant $\Delta E$ of about 170 meV and $\Delta E^{\text{fin}\to\text{ini}}$ of about 380 meV. As for the rings, this weak dependence on $Q$ indicates a local collapse process in which the topological charge is shed one unit at a time [Fig.~\ref{bags}(d)].

Again, we find two previously unreported collapse mechanisms in skyrmion bags, namely \textit{combining} collapse and \textit{kink} collapse, whose selection is determined by the parity of $Q$ [Fig.~\ref{bags}(a)]. For odd $Q$, two inner skyrmions merge, and the merged pair then annihilates, a pathway we term \textit{combining} collapse (circles in Fig.~\ref{bags}(a), $\text{BP}^{3\to2}$ in Fig.~\ref{bags}(a), (d); see Animation 2). The merged pair is locally a second-order skyrmion, and its collapse by injection of the opposite topological charge is known from frustrated magnets \cite{Desplat2019}. For even $Q$, an inner skyrmion couples to the outer boundary and triggers a local reversal of the topological charge density at a corner kink of the ring, which we term \textit{kink} collapse (squares in Fig.~\ref{bags}(a), $\text{BP}^{2\to1}$; see Animation 3). Kinks are topological excitations of the outer wall that carry part of the texture's charge \cite{Kuchkin2020}, making them the preferred sites for unwinding. The \textit{kink} pathway proceeds through an intermediate local maximum ($\mathrm{SP}^{*}$), where the inner skyrmion first overcomes its repulsion from the outer wall before reaching the true SP [Fig.~\ref{bags}(d)]. 
The odd-even selection rule discussed above originates from the orientation of the skyrmion bag relative to the atomic lattice.
As a result, the MEP obtained from the GNEB calculation depends on the initial orientation of the spin texture. The equilibrium orientation favors the collapse mechanisms described above, whereas rotating the spin texture before the MEP calculation can lead to a different pathway. For instance, a $Q=2$ bag can also collapse through the \textit{combining} mechanism. This interpretation is supported by additional GNEB calculations for rotated skyrmion bags (Supplementary Note 10 and Animation 4 in SM \cite{supplmat}). Since the corresponding energy barriers are of similar magnitude, this orientational dependence does not alter the physical conclusions of this work. We further explored a set of composite spin textures formed by combining high-$Q$ rings and bags. Despite their diverse initial spin structures, all configurations invariably relax to the same high-$Q$ skyrmion rings, uniquely determined by the total topological charge (Supplementary Note~9 in SM \cite{supplmat}). Overall, HOI stabilize the complete family of skyrmion bags, with collapse barriers that vary little with $Q$. Again, these predictions can be tested experimentally~\cite{Muckel2021,koraltan2025signatures}.

\section{SUMMARY AND PERSPECTIVE}
DMI has long been recognized as the key interaction responsible for noncollinear magnetic orders, including magnetic-field stabilized skyrmions. On its own, however, a homogeneous DMI generally stabilizes only $|Q|=1$ spin textures. Previous studies achieved high-$Q$ states using dipolar interactions, engineered DMI landscapes, or artificial confinement \cite{Niu2025,Kern2025,hassan2024dipolar,zhang2024stable}. In our work, we propose a novel stabilization mechanism driven by HOI for topological spin textures with arbitrary $Q$, including skyrmion/antiskyrmion rings as well as skyrmion bags. We further identify the recently proposed four-spin three-site interaction $Y_1$ \cite{Blugel2018,Hoffmann2020} as the key microscopic interaction underlying this stabilization. In MnPCl, the large negative $Y_1$ intrinsically favors spin canting, analogous to exchange frustration, thereby stabilizing high-$Q$ states even in the absence of DMI. Previous studies proposed that strong frustration of pairwise exchange interactions could stabilize skyrmions and antiskyrmions without DMI, but only for a specific ratio of competing exchange interactions \cite{leonov2015multiply}. In contrast, the HOI-based mechanism proposed here removes the need for fine-tuning. The mechanism is not limited to MnPCl. It extends to the broader family of Mn-based Janus monolayers, as we demonstrate for an additional material. Cr-based Janus monolayers, in contrast, are predicted to be much less promising (Supplementary Note~11 in SM \cite{supplmat}). The search for new hosts should therefore target a large negative $Y_1$ rather than a large DMI.

An important open question is how the energy barriers reported here set the thermal lifetimes of high-$Q$ topological spin textures.
Within harmonic transition state theory, the mean lifetime $\tau$ at temperature $T$ follows the Arrhenius law $\tau = \Gamma_0^{-1} \exp(\Delta E/k_{\text{B}}T)$. Here $\Delta E$ is the energy barrier of the transition and $\Gamma_0$ is the pre-exponential factor. The large barrier enhancement induced by the negative $Y_1$ is thus expected to increase the lifetime substantially. Quantifying it, however, also requires $\Gamma_0$, which has not been computed for high-$Q$ states with HOI. This is an important direction for future work and a key step toward room-temperature spintronic applications. The need is particularly pressing for skyrmion bags. They have already been created and stabilized experimentally, yet their thermal stability has never been quantified. More generally, the combination of arbitrary $Q$, compact size, and large energy barriers makes these states promising candidates for high-density spintronic devices. The high tunability of vdW magnets is a further asset. External stimuli such as strain or electric currents could switch between states of different topological charge \cite{Tang2021}. Because the emergent magnetic field scales with the topological charge, high-$Q$ states are also expected to exhibit enhanced topological Hall responses \cite{hassan2024dipolar}. We anticipate that our predictions can be
tested experimentally using SP-STM, which has already been demonstrated for skyrmions \cite{Muckel2021}. Our work also opens new perspectives for enhancing the stability of topological spin structures -- even in systems with inversion symmetry, \textit{i.e.}, in the absence of the DMI.

\section*{METHODS}\label{method_comp}

\subsection*{A. Magnetic interactions from first-principles calculations}

The magnetic interaction parameters in Eq.~(\ref{model}) are taken from the first-principles study of Janus monolayers by Li \textit{et al.}~\cite{Hongxin2023}. The corresponding spin spiral dispersions, shown in Supplementary Note~1 of SM \cite{supplmat}, illustrate the competition between the magnetic interactions and the resulting ground state.

Without HOI, flat spin spirals are exact single-\textbf{q} solutions of the Heisenberg exchange, DMI, and MAE Hamiltonians. The magnetic moment at lattice site $\mathbf{R}_i$ is described by
\begin{equation}
\mathbf{m}_i = 2 \left(\mathbf{R_q} \cos(\mathbf{q} \cdot \mathbf{R}_i) - \mathbf{I_q} \sin(\mathbf{q} \cdot \mathbf{R}_i)\right),
\label{eq:3}
\end{equation}
where $\mathbf{R_q}$ and $\mathbf{I_q}$ are orthogonal vectors of magnitude $1/2$. Substituting this ansatz into the Hamiltonian yields the spin-spiral dispersion, while the MAE contributes a constant energy shift of $K_u/2$ relative to the FM state.

With HOI, the degeneracy between single-\textbf{q} and multi-\textbf{q} states is lifted. The energy differences between three representative multi-\textbf{q} states (the two \textit{uudd} states and the $3Q$ state) and their corresponding single-\textbf{q} states are given, within the nearest-neighbor approximation, by~\cite{Hoffmann2020,paul2020role}
\begin{equation}
E_{\overline{\text{M}}}^{3Q} - E_{\overline{\text{M}}}^{1Q} = \frac{16}{3} \left( 2K_1 + B_1 - Y_1 \right),
\label{eq:4}
\end{equation}
\begin{equation}
E_{\overline{\Gamma \text{M}}/2}^{uudd} - E_{\overline{\Gamma \text{M}}/2}^{1Q} = 4 \left( 2K_1 - B_1 - Y_1 \right),
\label{eq:5}
\end{equation}
\begin{equation}
E_{3\overline{\Gamma \text{K}}/4}^{uudd} - E_{3\overline{\Gamma \text{K}}/4}^{1Q} = 4 \left( 2K_1 - B_1 + Y_1 \right).
\label{eq:6}
\end{equation}

Using the energies of both single-$\mathbf{q}$ and multi-$\mathbf{q}$ states, we determine the ground state of the system.

\subsection*{B. Initial state for solitons}

The initial spin configurations are defined on the discrete lattice by the unit magnetization field
$\mathbf{m}:\mathbb{R}^2\rightarrow\mathbb{S}^2$,
\begin{equation}\label{eq:skyrmion_ansatz}
    \mathbf{m}(\mathbf{r}) = \left( \begin{array}{c}    
    \cos\Phi(\mathbf{r})\sin\Theta(\mathbf{r})\\
    \sin\Phi(\mathbf{r})\sin\Theta(\mathbf{r})\\
    p\,\cos\Theta(\mathbf{r})
    \end{array}  \right),~ \Phi(\mathbf{r}) = \nu\phi(\mathbf{r})+\gamma ~.
\end{equation}
Here, $\phi(\mathbf{r})$ is the polar angle measured from the soliton center, $\nu\in\mathbb{Z}$ is the vorticity, $\gamma\in[0,2\pi)$ is the helicity, and $p=\pm1$ is the polarity of the core. The radial profile is defined as
\begin{equation}\label{eq:theta}
    \Theta(r)=\pi-\arcsin\!\left[\tanh\frac{2(r+c)}{w}\right]-\arcsin\!\left[\tanh\frac{2(r-c)}{w}\right],
\end{equation}
with $\Theta(0)=\pi$ and $\Theta(r\!\rightarrow\!\infty)=0$. It crosses $\Theta=\pi/2$ at $r\simeq c$, so $c$ is the radius of the reversed core and $w$ the width of the wall that separates it from the collinear background. Inserting Eq.~(\ref{eq:skyrmion_ansatz}) into Eq.~(\ref{topo_charge}) gives
$Q=-p\nu$. Throughout this work, the background magnetization is aligned with the applied field ($p=+1$), so $Q=-\nu$ for any texture obtained directly from Eq.~(\ref{eq:skyrmion_ansatz}). For $\nu=\pm1$ this gives a skyrmion and an antiskyrmion, respectively.

High-$Q$ rings follow from Eq.~(\ref{eq:skyrmion_ansatz}) by raising the vorticity to $|\nu|=|Q|>1$. 
Upon relaxation, the initially reversed core evolves into a ring-shaped domain wall. Skyrmion bags cannot be obtained from a single evaluation of Eq.~(\ref{eq:skyrmion_ansatz}) and are constructed in two steps. We first initialize a single skyrmion ($\nu=+1$, $p=+1$) over the entire lattice, whose core points opposite to the applied field. We then embed $n$ smaller skyrmions with opposite polarity ($\nu=+1$, $p=-1$), equally spaced on a circle within the reversed core, so that their cores are aligned with the field. Reversing the polarity also reverses the domain-wall rotation. To keep the sense favored by the DMI, the in-plane components of the inner skyrmions point inward while those of the outer skyrmion point outward. All textures remain N\'eel-type, although the inner and outer skyrmions differ in helicity by $\Delta\gamma=\pi$. Each inner skyrmion is written only inside a restricted region, leaving the domain wall of the outer skyrmion unchanged.

The inner skyrmions do not overlap, so their topological charges add to that of the outer skyrmion, giving $Q=n-1$, where $n$ is the number of inner skyrmions. We consider $n=1$--$5$, corresponding to $Q=0$--$4$, the case $n=0$ recovers the isolated skyrmion. 
The special case $n=1$ ($Q=0$) is the skyrmionium, in which the inner and outer skyrmions are concentric. Composite ring--bag states (Supplementary Note~9 in SM \cite{supplmat}) are constructed using the same procedure, except that the outer skyrmion is replaced by a high-$Q$ ring initialized with $|\nu|>1$.

\subsection*{C. Equilibrium state calculation}

Metastable spin textures are local minima of the Hamiltonian in Eq.~(\ref{model}). We construct initial spin textures such as skyrmions, antiskyrmions, high-$Q$ skyrmion rings, or skyrmion bags within the field-polarized FM background and relax them to local minima using the velocity projection optimization (VPO) algorithm~\cite{bessarab2015method}, as implemented in the \textsc{spinaker} code. Throughout the relaxation, $Q$ is conserved for all textures studied in this work. The relaxation is considered to be converged when the maximum torque on any spin drops below $10^{-12}$~eV. All calculations are performed on $120\times120$ spin lattices with periodic boundary conditions. The fully relaxed spin textures then serve as the initial states for the subsequent GNEB calculations.

\subsection*{D. Topological charge density calculation}

To characterize magnetic solitons, we calculate the topological charge $Q$ as the sum of the topological charge density $q$ over selected spin textures. For a discrete model, the topological charge in Eq.~(\ref{topo_charge}) from the main text is evaluated using the lattice discretization formulated by Berg and L\"uscher~\cite{Berg1981},
\begin{equation}
Q=\sum_{\mathbf{r}^*} q(\mathbf{r}^*)~,
\end{equation}
where the topological charge density $q(\mathbf{r}^*)$ is defined as a function of the four magnetic moments $\mathbf{m}_1$, $\mathbf{m}_2$, $\mathbf{m}_3$, and $\mathbf{m}_4$ at the lattice sites closest to $\mathbf{r}^*$ on the triangular lattice,
\begin{equation}
q(\mathbf{r}^*)=
\frac{1}{4\pi}
\left[
\mathcal{A}(\mathbf{m}_1,\mathbf{m}_2,\mathbf{m}_3)
+
\mathcal{A}(\mathbf{m}_1,\mathbf{m}_3,\mathbf{m}_4)
\right]~.
\end{equation}
Here, $\mathcal{A}(\mathbf{m}_1,\mathbf{m}_2,\mathbf{m}_3)$ denotes the signed area of the spherical triangle with corners $\mathbf{m}_1$, $\mathbf{m}_2$, and $\mathbf{m}_3$.

\subsection*{E. MEPs and saddle point search}

Energy barriers are obtained from MEPs calculated with the GNEB method~\cite{bessarab2015method}. A path between two magnetic states is represented by a chain of $N$ images, whose endpoints are the relaxed states obtained by minimization. The intermediate images are generated via geodesic interpolation with a small random perturbation, thereby breaking the symmetry of the initial path. Each image then relaxes under the force
\begin{equation}
\mathbf{F}_i=-\left.\nabla E(\mathbf{M}_i)\right|_{\perp}
+\left.\mathbf{F}_i^{\mathrm{spring}}\right|_{\parallel}~,
\label{eq:gneb}
\end{equation}
where $\mathbf{M}_i$ is the spin configuration of image $i$. 
Once the path is pre-converged, the highest-energy image is switched to the climbing-image scheme (CI-GNEB),
\begin{equation}
\mathbf{F}_i^{\text{CI}}=-\left.\nabla E(\mathbf{M}_i)\right|_{\perp} +\left.\nabla E(\mathbf{M}_i)\right|_{\parallel},
\label{eq:cigneb}
\end{equation}
in which the spring force is removed, and the parallel gradient component is inverted. This image moves uphill along the path and converges onto the first-order saddle point (SP). The calculation stops when the maximum geodesic force drops below $10^{-8}$~eV/rad. The resulting energy barrier is $\Delta E=E_{\text{SP}}-E_{\text{ini}}$ and the nucleation barrier is $\Delta E^{\text{fin}\rightarrow\text{ini}}=E_{\text{SP}}-E_{\text{fin}}$, with $E_{\text{ini}}$ and $E_{\text{fin}}$ the energies of the initial and final states of the transition.

For the high-$Q$ skyrmion rings, the collapse proceeds through a sequence of transitions whose final states are not known \textit{a priori}. We therefore locate first-order SP using the geodesic minimum mode following (GMMF) method~\cite{Muller2018,Schrautzer2025}. Unlike GNEB, GMMF converges directly to a first-order SP from a single spin configuration, without an initial path or a predefined final state. The 
component of the energy gradient along the minimum mode $\textbf{q}=(\textbf{q}_1,\ldots,\textbf{q}_N)$, the Hessian eigenvector with the lowest eigenvalue, is inverted, which yields the modified effective field
\begin{equation}
\textbf{b}_i=-\boldsymbol{\nabla}_i E
+2\,(\boldsymbol{\nabla}E\cdot\textbf{q})\,\textbf{q}_i~,
\label{gmmf_field}
\end{equation}
under which the first-order SP behaves like a minimum. Projecting $\textbf{b}_i$ onto the local tangent space gives the GMMF force
\begin{equation}
\textbf{f}_i=\textbf{b}_i-(\textbf{b}_i\cdot\textbf{m}_i)\,\textbf{m}_i~.
\label{gmmf_force}
\end{equation}
under which the spin configurations are relaxed. The minimum mode is obtained at each iteration from the generalized Rayleigh quotient with an L-BFGS algorithm, without constructing the Hessian explicitly~\cite{Schrautzer2025}. The calculation is converged when the maximum force drops below $10^{-12}$~eV, and the configuration is accepted as a first-order SP when the second-lowest Hessian eigenvalue is positive. When a GNEB path converges, GMMF calculations are initialized from the highest-energy image, and the resulting energy barriers agree with those obtained from CI-GNEB. 

High-$Q$ rings and bags decay through a sequence of elementary transitions, each reducing $|Q|$ by one. The MEPs for high-$Q$ states are therefore obtained by calculating each elementary transition separately and concatenating the resulting paths. A discontinuous change in the topological charge $Q$ between neighboring images identifies the Bloch point (BP), whereas the corresponding topological charge density $q(\mathbf{r}^*)$ at the BP distinguishes the underlying collapse mechanism. All GNEB calculations are performed with 60 images. Both GNEB and GMMF calculations are carried out with the \textsc{spinaker} code.\\

\section*{REFERENCES}
\bibliography{References}

\hfill\break

\section*{ACKNOWLEDGMENTS} 

This work is supported by France 2030 government investment plan managed by the French National Research Agency under grant reference PEPR SPIN – [SPINTHEORY] ANR-22-EXSP-0009, the National Natural Science Foundation of China (Grant No.11804301), the Natural Science Foundation of Zhejiang Province (Grant No.LMS25A040001), and the Funds of the Natural Science Foundation of Hangzhou (Grant No.2025SZRJJ0830). This study has been (partially) supported through the grant NanoX no.~ANR-17-EURE-0009 in the framework of the ``Programme des Investissements d’Avenir". This work was performed using HPC resources from CALMIP (Grant 2024/2026-[P21008]).

\emph{}

\section*{AUTHOR CONTRIBUTIONS}

D.L. initiated and supervised the project. Z.Su. performed the energy minimizations and GNEB/GMMF simulations with the help of S.Zh., M.A.G., and D.L. M.Ar. worked on spin spiral dispersions. Z.Su. is supervised by C.S. and D.L. Z.Su. prepared the figures based on suggestions from S.Zh., M.A.G., and D.L. All authors contributed to the analysis and discussion of the simulation results. Z.Su. and D.L. wrote the paper, and all authors contributed to the revision of the manuscript.

\emph{}

The authors declare no competing interests.

\emph{}

Correspondence and requests for materials should be addressed to Dongzhe Li.\\

\section*{DATA AVAILABILITY}
All the data are available from the corresponding authors upon reasonable request. Source data are provided with this paper. 

\end{document}